\documentclass[11pt]{amsart}
\usepackage{geometry}                % See geometry.pdf to learn the layout options. There are lots.
\usepackage{graphicx}
\usepackage{epstopdf, epsfig}
\usepackage{graphicx,epsfig}
\usepackage{amsmath}
\usepackage{amsfonts}
\usepackage{amssymb}
\usepackage{caption}
\usepackage{subcaption}
\usepackage{color}
\graphicspath{{./figures/}{/home/guy/Dropbox/ve_chapter/Soft_swimmers_Feb2016/figures/}{/homes/home03/g/guy/Dropbox/ve_chapter/Soft_swimmers_Feb2016/figures/}}

\definecolor{review_color}{rgb}{0.0, 0.0, 0.0}  % set review color to black
\definecolor{review_color2}{rgb}{0.0, 0.0, 0.0} 
\definecolor{review_color3}{rgb}{0.0, 0.0, 0.0}

\def\fun#1#2{\lower3.6pt\vbox{\baselineskip0pt\lineskip.9pt
  \ialign{$\mathsurround=0pt#1\hfil##\hfil$\crcr#2\crcr\sim\crcr}}}
\newcommand{\ip}[1]{\left<#1\right>}

\newcommand{\bu}{{\bf u}}

\newcommand{\F}{\mathbf{F}}

\newcommand{\eps}{\varepsilon}

\newcommand{\f}{\mathbf{f}}

\newcommand{\lap}{\triangle}

\renewcommand{\d}{\partial}
\renewcommand{\div}{\nabla\cdot}

\newcommand{\dgrad}{\cdot\nabla}

\renewcommand{\u}{\textbf{u}}

\def\lf{\left}
\def\rt{\right}

\newcommand\X{\mathbf{X}}

\newcommand{\G}{\textit{G}}

\newcommand{\taup}{\boldsymbol{\tau_p}}

\newcommand{\kb}{\textit{B}}
\newcommand{\kbdim}{\mathcal{B}}
\newcommand{\zetave}{\zeta_{\textrm{ve}}}

\newcommand{\De}{\textrm{De}}     % Deborah    number
\newcommand{\strDe}{\textrm{Stroke}_\textrm{De}}     % Deborah    number
\newcommand{\per}{T}
\newcommand{\headamp}{A_h}
\newcommand{\tailamp}{A_t}
\newcommand{\dgamma}{\dot{\boldsymbol{\gamma}}}

\newcommand{\ff}{{f}_{\textrm{fluid}}}
\newcommand{\fv}{{f}_{\textrm{vis}}}

\usepackage{bm}

\title[Flexible Swimmers]{The role of body flexibility in stroke enhancements for finite-length undulatory swimmers in viscoelastic fluids}
\author{Becca Thomases and Robert D. Guy}

\begin{document}

\maketitle

\begin{abstract}
The role of passive body dynamics on the kinematics of swimming micro-organisms in complex fluids is investigated. Asymptotic analysis of small amplitude motions of a finite-length undulatory swimmer in a Stokes-Oldroyd-B fluid is used to predict shape changes that result as body elasticity and fluid elasticity are varied. Results from the analysis are compared with numerical simulations, 
 and the numerically simulated shape changes agree with the analysis at both small and large amplitudes, even for strongly elastic flows. We compute a stroke-induced swimming speed that accounts for the shape changes, but not additional effects of fluid elasticity. Elasticity-induced shape changes lead to larger amplitude strokes for sufficiently soft swimmers in a viscoelastic fluid, and these stroke boosts can lead to swimming speed-ups. However, for the strokes we examine, we find that additional effects of fluid elasticity generically result in a slow-down. Our high amplitude strokes in strongly elastic flows lead to a qualitatively different regime in which highly concentrated elastic stresses accumulate near swimmer bodies and dramatic slow-downs are seen. 

\end{abstract}

%\begin{keywords}
%\end{keywords}

% \section{Introduction}
 
\section{Introduction}

%Complex fluids are everywhere, micro-organism swimming is important.
%
There has been an intense effort over the past 10 years to understand
the effect of fluid elasticity on micro-organism
swimming. Experiments, analysis, and simulations of low-Reynolds
number swimming of microorganisms in complex fluids, in particular
viscoelastic fluids, has led to a variety of results \--- some
complimentary, some apparently conflicting \--- on the effect of fluid
elasticity on swimming speed.  We know that gait, body
stiffness, and nonlinear effects matter, but we still do not have a
clear understanding of how they interact during locomotion.

%%%%%%%%%%%
\textcolor{review_color2}{
Early work quantifying the effect of fluid elasticity on swimming using a linear constitutive law for the fluid and asymptotic analysis of small amplitude motions showed that elasticity had no effect on swimming speed but increased swimming efficiency \cite{chaudhury1979swimming,sturges1981motion}. However, in \cite{LAUGA:2007} a full analysis of the classical Taylor swimming sheet for small amplitude undulatory motion showed that the nonlinearities of the viscoelastic fluid model must be included in a computation of swimming speed, and found that swimming speed is always hindered by fluid elasticity. Similar small amplitude asymptotic analysis was done for waving filaments and helices \cite{fu2007theory, fu2009swimming} also predicting slow-downs due to fluid elasticity.  \cite{elfring2016effect} and \cite{riley2015small} have demonstrated the importance of the details of the swimming gait in understanding the effect of fluid elasticity on swimming speed, indeed showing that elastic speed-ups are possible for some gaits, which is further highlighted by the analysis of three-sphere swimmers \cite{curtis2013three}.}

\textcolor{review_color2}{Biological swimmers have been shown to change their gait in response to changes in rheology \cite{shen2011undulatory,gagnon2014undulatory,qin2015flagellar}, making it hard to interpret the mechanisms responsible for observed changes in swimming performance. In more controlled physical models of swimmers in different fluids 
a variety of results have shown that fluid elasticity can boost swimming speed \cite{liu2011force,keim2012fluid,espinosa2013fluid} or retard swimming speed \cite{dasgupta2013speed,godinez2015complex}. Swimmers with large amplitude motions have
been theoretically investigated using numerical simulations, and have added significant information about the response of swimmers to fluid elasticity with a variety of swimming gaits \cite{teran2010viscoelastic,balmforth2010microelastohydrodynamics, zhu2012self,spagnolie2013locomotion,montenegro2013physics,Thomases2014d,li2014effect,li2015undulatory,salazar2016numerical}.  \textcolor{review_color3}{In addition, for recent reviews of 
swimming in complex fluids, see \cite{elfring2015theory} for a theoretical view, and see \cite{sznitman2015locomotion} for  
an experimental view. }
These many studies have focused on different types of swimmers, in different fluid rheologies, and despite the wealth of results we still lack an understanding of the underlying principles of swimming in complex fluids. }

%%%%%%%%%%%%

To try to isolate
physical mechanisms that are significant in a variety of biologically relevant problems, but simple enough to analyse,
we focus here on undulatory swimmers in an Oldroyd-B fluid. Even in this more restrictive setting, we nevertheless
still find apparently contradictory results and a lack of mechanistic explanations for those differences. 
Asymptotic analysis of infinitely long, prescribed shape, small-amplitude swimmers has
shown that swimming is hindered by the addition of elastic stresses
\cite{LAUGA:2007}, although allowing for flexibility can lead to
enhancements \cite{riley2014enhanced}.  Biological experiments have
shown a viscoelastic slow-down for \textit{C.\ elegans}
\cite{shen2011undulatory}, while simulations of finite-length
swimmers with large tail amplitudes
\cite{teran2010viscoelastic,Thomases2014d} give a
non-monotonic boost as fluid elasticity is varied.  In
\cite{Thomases2014d} we concluded that shape changes due to body
flexibility and fluid elasticity are important, but those results
did not explain the results from a physical experiment which
showed monotonic speed-ups due to fluid elasticity in swimmers
with large tail amplitudes \cite{espinosa2013fluid}. Furthermore, recent numerical
simulations \cite{salazar2016numerical} appear to contradict the speed-ups reported in \cite{teran2010viscoelastic,Thomases2014d}.

 \textcolor{review_color3}{The relevance of body elasticity in viscoelastic speed enhancements was identified for small amplitude infinite length swimmers in \cite{riley2014enhanced}, where the authors attribute the speed enhancements to a viscoelastic ``suction" which results in an amplitude boost.  However, their analysis does not extend to finite-length large amplitude swimmers where  
the role of elasticity-induced shape changes has not been addressed directly.}
%
%No one has visited for finite length, 
%
%the analyis dosent extend to finite length or large amplitude
%
%At small amplitude  length changes are higher order effects (cite taylor)  .  at large amplitude it isn't known?  
%
%Their results are similar to what we find, but not generalizable directly to  large amplitude, finite-length swimmers. }
%%In that study the swimmer body was infinite length and extensible, hence a boost in amplitude of a single frequency wave necessarily changes its length.  Such length changes cannot occur for finite-length inextensible swimmers.}
%
%
The disparity of the results in \cite{teran2010viscoelastic,Thomases2014d,espinosa2013fluid,salazar2016numerical},
 all focusing on large amplitude, finite-length, undulatory swimmers in Oldroyd-B fluids, indicates that something is missing in our understanding of the problem.

There remains a gap between our understanding from analysis and what
we see in computational, biological, and physical experiments. Here we
combine analysis with numerical simulations of finite-length large
amplitude swimmers to show how fluid elasticity induces shape changes
in finite-length flexible swimmers and how those shape changes can lead to speed
boosts.  We show how shape changes depend on both body stiffness
as well as fluid elasticity and analyse the effect that shape
changes alone have on swimming speed.

%%%%%%%%%%%%%%%%%%%%%%%%%%%%%%%%%%%%%%%%%%%%%%%%%%%%%%%%%%%%%%%%%%%%%%%%%%%
%
% Section 2
%
%%%%%%%%%%%%%%%%%%%%%%%%%%%%%%%%%%%%%%%%%%%%%%%%%%%%%%%%%%%%%%%%%%%%%%%%%%%

\section{Effect of passive body dynamics}\label{model}

\subsection{Methodology} We follow the computational framework in
\cite{Thomases2014d,guy2015computational}, where the swimmer is modeled as
an inextensible flexible sheet of finite-length $L$ immersed in a 2D
fluid.  We describe the undulatory motion of the swimmer by a curvature of the form
\begin{equation}
  \kappa_0(s,t)=\left(\tailamp(L-s)/L+\headamp s/L\right)
                \sin(2\pi t/\per+\pi s),
  \label{eq_curve}
\end{equation}
where $s\in[0,L]$ is the body coordinate. Here $\tailamp$ is the
curvature amplitude at the ``tail" ($s=0$) and $\headamp$ is the
curvature amplitude at the ``head" ($s=L$) of the swimmer.

We use the immersed boundary method to solve for the coupled motion of
the fluid and the swimmer \cite{fauci1988computational}.  Both
inextensibility and shape are imposed (approximately) by forces that
are designed to penalize extension and deviation from the prescribed
target curvature.  These forces are derived from the variation of
bending and extension (stretching) energy functionals.  For example,
the bending energy is
\begin{equation}
  E_{\rm b} = \kb/2\int_{\Gamma}\lf(\kappa-\kappa_{0}\rt)^{2}\,ds,
  \label{bending_energy:eq}
\end{equation}
where $\kb$ is the bending
stiffness, $\kappa$ is the curvature of the sheet, and $\kappa_{0}$ is
the prescribed target curvature. One can interpret the model as an
active sheet with bending stiffness $\kb$ driven by an active body
moment density $\kb\kappa_0$. We scale forces relative to viscous
forces so that for $\kb\gg 1$, the realized shape of the swimmer is
very close to the prescribed shape.  For $\kb\sim 1$, the realized
shape is the result of fluid-structure interaction; i.e.\ passive body dynamics
influence the resulting stroke. 

The viscoelastic fluid is described by the Oldroyd-B model at zero
Reynolds number \cite{BHAC1980}, regularized by stress diffusion \cite{SB1995,Thomases2011}.  
The system of equations describing the fluid are
\begin{gather}
 \Delta \u - \nabla p +\xi \nabla\cdot\taup + \f = 0,\label{mombaleq}\\ \div \u=0,\label{consmasseq} \\
 % \De\stackrel{\nabla}\taup+\taup=\boldsymbol{\dot{\gamma}}+\De\;\varepsilon\lap\taup\label{stresseq}
  \De\bigl({\d\taup}/{\d t}+\u\dgrad\taup 
           -\nabla\bu~\taup-\taup~\nabla\bu^T 
      \bigr)
  +\taup=\boldsymbol{\dot{\gamma}}+\De\;\varepsilon\lap\taup\label{stresseq}
\end{gather}
where $\u$ is the fluid velocity, $p$ is the pressure, $\taup$ is the
viscoelastic stress, $\boldsymbol{\dot{\gamma}}$ is the rate of strain
tensor, and $\f$ is the elastic force density generated by the
swimmer. 
%%
%% The upper convected time derivative is defined by
%% $\stackrel{\nabla}{\taup}\equiv {\d\taup}/{\d t}+\u\dgrad\taup -\left(
%% \nabla\bu~\taup+\taup~\nabla\bu^T \right)$.  
%%
 Here $\xi$
is the polymer to solvent viscosity ratio, $\De=\lambda/\per,$  the Deborah number is
the ratio of elastic relaxation time to stroke period, and $\varepsilon\ll 1$ is the stress diffusion
coefficient.
%
%We non-dimensionalize with a characteristic length scale of $1$
%mm, time scale of $\per=1$ s, and stress scale $\mu/\per$, where
%$\mu$ is the viscosity of water.  
%
%
%Other model and numerical parameters
%are given here \footnote{
%

The system is solved in a 2D periodic domain of size
$[0,2]\times[0,2],$ with $L=1,$ $dt=10^{-3},$ and $dx=2^{-8}.$ We fix
$\xi=0.5,$ consistent with \cite{teran2010viscoelastic}, and
$\eps=0.0015$ which provides a regularization to
control large stress gradient growth
\cite{Thomases2011}. We enforce
inextensibility with a dimensionless stiffness constant of $2500$.
%}.

%
%%%%%%%%%%%%%%%%%%%%%%%%%%%%%%%%%%%%%%%%%%%%%%%%%%%%%%%%%%%%%%%%%%%%%%%%%%%

\subsection{Varying body stiffness}

\begin{figure}
\centering
\includegraphics[width=.95\textwidth]{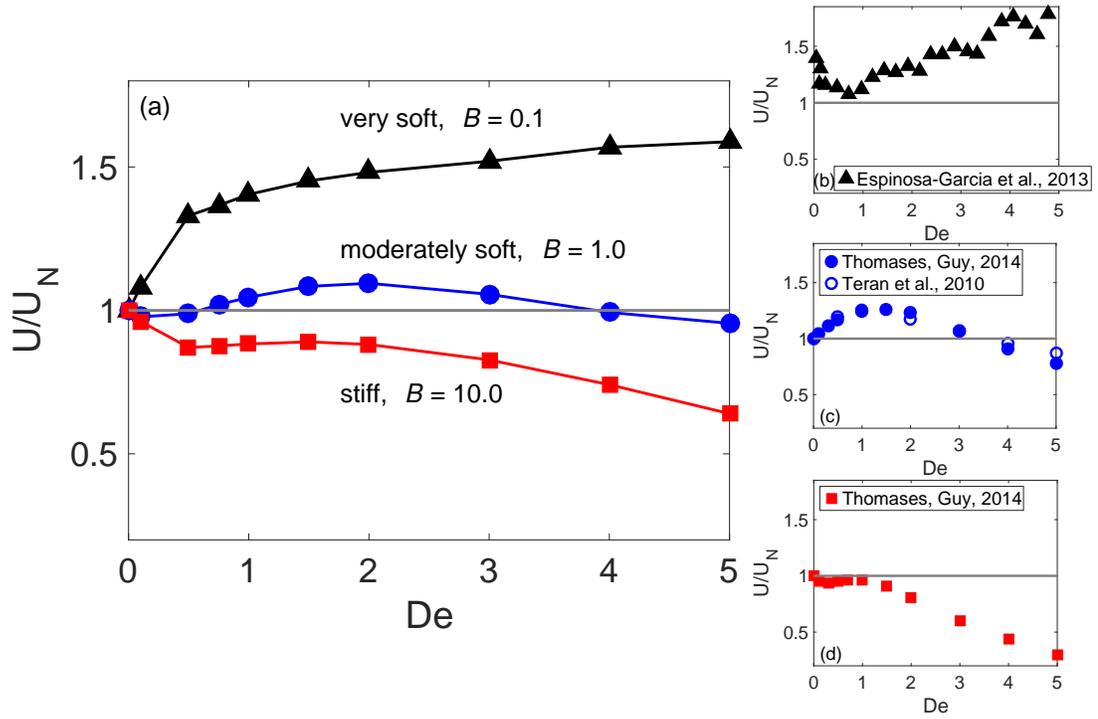}\caption{(a) Swimming speed (normalized by Newtonian swimming speed) as a function
  of $\De$ for different bending stiffness $\kb.$  Here we fix the period, 
  $T=1$. (b)-(d) Normalized swimming speed as a function of $\De.$ Reproductions from the literature: (b) From \cite{espinosa2013fluid}, 
  a physical model of a swimmer with a flexible tail. (c) From \cite{teran2010viscoelastic}, and \cite{Thomases2014d},  
   two different numerical simulations for a soft stroke with a large amplitude tail (d) From \cite{Thomases2014d}, numerical simulations for a stiff stroke with a large amplitude tail. }
\label{fig_vg}
\end{figure}

To understand the role of body elasticity, we use our simulations to
calculate the Stokes-normalized swimming speed while varying $\kb$ and
$\De$ for a fixed period ($\per=1$).  We use a stroke defined by equation
\eqref{eq_curve} with $A_t=5,$ and $A_h=2.$ This gives a
high-amplitude stroke like in
\cite{teran2010viscoelastic,Thomases2014d}.  In figure
\ref{fig_vg}(a) we plot normalized swimming speed as a function of
$\De$ for three characteristic stiffness values of $\kb=0.1,1.0,10.0$,
which we refer to as \textit{very soft, moderately soft}, and
\textit{stiff}, respectively. For \textit{very soft} swimmers we see a
monotonic boost in swimming speed, with a greater than 50$\%$ boost 
for high $\De.$

%\begin{figure}
%\centering
%\includegraphics[width=.95\textwidth]{figure1.eps}\caption{(a) Normalized swimming speed as a function
%  of $\De$ for different bending stiffness $\kb,$ with 
%  $T=1$. (b)-(d) reproductions of data from
%  literature.}
%\label{fig_vg}
%\end{figure} 

This response is similar to what was reported in
\cite{espinosa2013fluid} using a physical model of a swimmer with a
flexible tail (figure \ref{fig_vg}(b)). For \textit{moderately soft}
swimmers, we see a non-monotonic speed-up, including a smaller speed
boost over the Newtonian speed, followed by a slow-down at larger
$\De.$ This type of non-monotonic speed-up was first reported in
\cite{teran2010viscoelastic} and again in \cite{Thomases2014d} for a
soft stroke with high amplitude (figure \ref{fig_vg}(c)). Finally, for
\textit{stiff} swimmers we see non-monotonic behavior but no boost
over the Newtonian speed, again followed by a slow-down at larger
$\De.$ This type of slow-down was reported in \cite{Thomases2014d}
for a stiff kicker (figure
\ref{fig_vg}(d)).

\begin{figure}
\centering
\includegraphics[width=.5\textwidth]{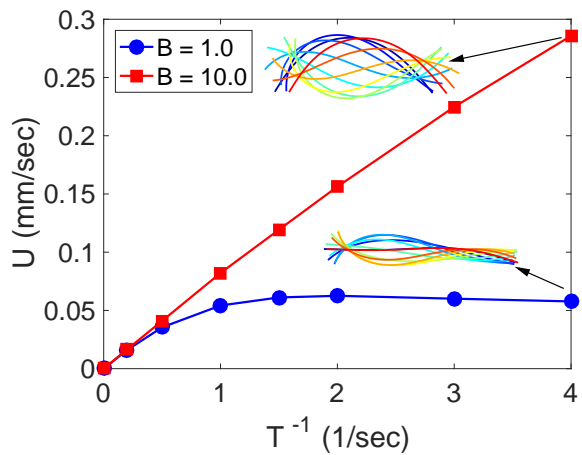}
\caption{\textcolor{review_color}{Dimensional swimming speed in a Newtonian fluid for \textit{stiff,} $\kb=10.0,$ and \textit{moderately soft,} $\kb=1.0,$ swimmers over a range of frequencies $\per^{-1}.$ Inset figures show shapes of swimmer over a period at the highest computed frequency for both soft and stiff swimmers. }}\label{fig_vss}
\end{figure}

In contrast to stiff, or rigid, swimmers, the dynamics of flexible
swimmers involve an additional time scale.  In a viscous fluid, rigid
swimmers move with a velocity proportional to the beat frequency (the
only time scale in the problem). 
The problem of a rigid swimmer in a viscoelastic fluid has two time
scales, the beat frequency and the relaxation time, whose ratio 
is the dimensionless relaxation time $\De.$ 
The swimming speed of soft swimmers
depends nonlinearly on the frequency because the shape changes with
the frequency.  \textcolor{review_color}{Figure \ref{fig_vss} shows the swimming speed in a Newtonian
fluid for both a \textit{stiff} swimmer,  $\kb=10.0$ and a \textit{moderately soft} swimmer $\kb=1.0$ over a range
of beating frequencies. We see a linear response in the case of a \textit{stiff} swimmer and a nonlinear response for the \textit{moderately soft} swimmer.  Inset swimmer shapes show how the shape changes as a function of the stiffness in the high frequency case. }

\begin{figure}
\centering
\includegraphics[width=.95\textwidth]{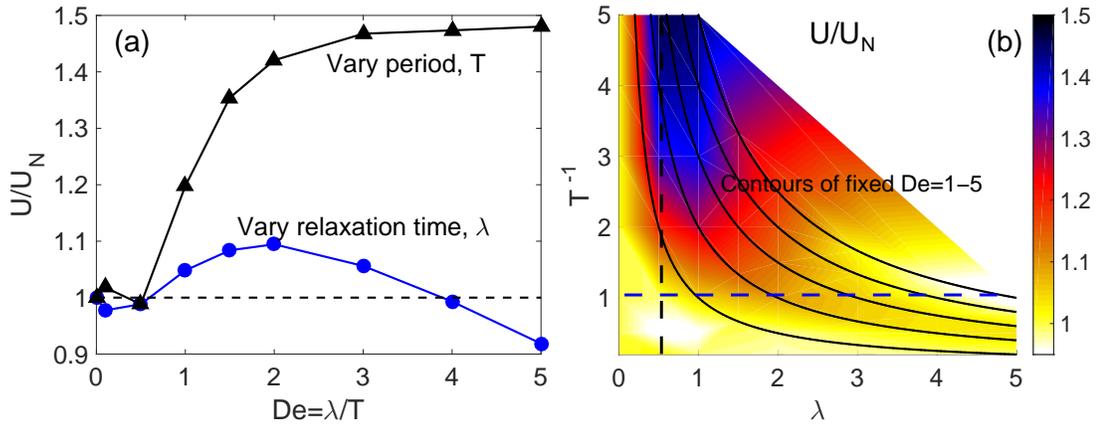}
\caption{(a) Normalized swimming speed as a function of $\De.$ Curves generated by varying only 
relaxation time (for fixed period $\per=1$) or stroke period (for fixed relaxation time $\lambda=0.5$). Body stiffness is fixed:
$\kb=1.0$. \textcolor{review_color}{(b) Normalized swimming speed as a function of both $\lambda$ and $T,$ with contours overlayed for constant $\De$ values. Dashed lines correspond to the locations of the data in \ref{fig_samede} (a). }
%
%Other relevant stroke parameters: $A_t = 5, A_h = 2$ and $\kb=1,$ see section \ref{model}.  
%
}\label{fig_samede}
\end{figure}

To illustrate the significance of multiple time scales for flexible
swimmers in viscoelastic fluids
we compute the Stokes-normalized swimming speed as a function of $\De$
varied two ways: by varying the relaxation time for a fixed
period, and by varying the period for a fixed relaxation
time. Both simulations are performed with the same bending stiffness,
$\kb=1.0$, where passive body dynamics are significant. Results are
shown in figure \ref{fig_samede} (a) for a swimmer with the same stroke
from figure \ref{fig_vg}, and the two curves show remarkable
qualitative differences.  
For a rigid swimmer these would give
equivalent results. 
 Thus
this third time-scale, arising from body flexibility, needs to be explicitly included in any discussion of swimming in elastic fluids.  \textcolor{review_color}{A more complete picture
of how the swimming speed depends on both the relaxation time and period when the body is soft, is shown in figure \ref{fig_samede} (b).
%, where a plot of the normalized swimming speed as a function of both $\lambda$ and $\per^{-1}$ is shown for the same data as in figure \ref{fig_samede} (a).
Contours of constant $\De=1-5$ are overlayed in black and the effect of body stiffness is clearly evident as you see the swimming speed vary significantly along any of the contours. The dashed lines denote the locations of the data in figure \ref{fig_samede} (a).}

%%%%%%%%%%%%%%%%%%%%%%%%%%%%%%%%%%%%%%%%%%%%%%%%%%%%%%%%%%%%%%%%%%%%%%%%%%%
%
% Section 3
%
%%%%%%%%%%%%%%%%%%%%%%%%%%%%%%%%%%%%%%%%%%%%%%%%%%%%%%%%%%%%%%%%%%%%%%%%%%%

\section{Analysis of shape changes}\label{Sec_shape}
The effect of body stiffness on swimming kinematics has been
previously studied for viscous fluids
\cite{wiggins1998flexive,lauga2007floppy}. Shape changes in viscoelastic fluids have been
examined \cite{fu2008beating}, but the relationship between shape
changes and swimming speed has not been examined for finite-length
swimmers.  Here we review the theory and compare it with numerical
simulations.

\subsection{Linear theory: Newtonian fluids} \label{flex_visc}
We begin by considering small amplitude displacements of a finite-length
thin elastic rod in a  Newtonian fluid driven by prescribed curvature,
$\kappa_{0}(s,t)$, (equivalently, prescribed moments) along the body
with free ends.  \textcolor{review_color}{The theoretical analysis
is similar in 2D and 3D, however we will make note of the differences when we compare with the numerical simulations in 2D. We proceed with the analysis in 3D for simplicity.}  The shape of the rod is determined by the
balance between elastic forces and viscous drag. The vertical
displacement, $y(s,t)$ satisfies
\begin{gather}\label{beameq}
  \zeta_{\perp}y_t=-\kbdim(y_{ss}-\kappa_0)_{ss},\\
   y_{ss}-\kappa_0=0, (y_{ss}-\kappa_0)_s=0, \textrm{ at } s=0,L.
\end{gather}
Here $\zeta_{\perp}$ is the perpendicular drag coefficient and $\kbdim$
is the bending stiffness of the rod. \textcolor{review_color}{Note that the use of new notation $\kbdim,$ is intended to distinguish this (dimensional) bending stiffness we use in the linear theory from our previously defined (non-dimensional) bending stiffness, $\kb$ which we use in our numerical simulations. }

%
% We can interpret $\kb(\kappa_0)_{ss}$ as an active driving moment
% prescribed along the body.
% While this method of driving the swimmer is
% different from prescribing a driving force or torque at the ends, the
% analysis and results from the analysis are quite similar.
%

Nondimensionalizing equation \eqref{beameq} using the body length $L$
and the period of the driving force $T$ results in the dimensionless
parameter we call the body response time:
\begin{equation}\label{Gdef}
  \G=\frac{T}{\kbdim^{-1}\zeta_{\perp}L^4}=
  \frac{\textrm{period of motion}}
       {\textrm{elasto-hydrodynamic beam relaxation time}}.
\end{equation} We note that $\G^{-1}$ could be called a body relaxation
time.
The same nondimensional group has appeared previously, but has been
interpreted differently. In \cite{shelley2000stokesian} a quantity
similar to $\G$ was considered an ``effective viscosity'' of growing
elastic filaments. The Sperm number ($Sp=\G^{-1/4}$) is the ratio of
the body length to the viscous decay length
\cite{wiggins1998flexive,fu2007theory}.  This interpretation is natural when
considering filaments driven at one end rather than along the body as
we do here.

We change variables from displacement to curvature deviation
\[c(s,t)=\kappa(s,t)-\kappa_0(s,t)\] to facilitate comparing with large
amplitude simulations. For small displacements
$y_{ss}(s,t)\approx \kappa(s,t)$, and equation \eqref{beameq},
(non-dimensionalized) becomes
\begin{gather}\label{curveq}
  c_t  =-\G c_{ssss}-\frac{\d\kappa_0}{\d t} \\\label{bc}
  c=0, c_s=0 \textrm{ at } s=0,1.
\end{gather}
%$ c_t  =-\G c_{ssss}-\frac{\d\kappa_0}{\d t},$ with boundary conditions, $c=0, c_s=0 \textrm{ at } s=0,L.$
%\end{gather}
For a given $\kappa_0,$ we use an orthogonal function expansion to
solve the non-dimensional equations for $c(s,t)$.  We let the driving curvature be given as 
%  $\kappa_0(s,t)=\sum_{k=1}^{\infty} \alpha_k^{\infty} e^{2\pi i k t}\Psi_k(s),$
%
\begin{equation}\label{kappa0}
  \kappa_0(s,t)=\sum_{k=1}^{\infty} \alpha_k^{\infty} e^{2\pi i \mu_k t}\Psi_k(s),
\end{equation}
and solve the eigenvalue problem,  \begin{gather*}\label{evprob}\mu \Psi(s)=-\Psi_{ssss},\\\label{ev_bc} \Psi=0, \Psi_s=0 \textrm{ at } s=0,1,\end{gather*} for eigenvalues $\mu_k$ and eigenfunctions $\Psi_k(s).$  The expansion coefficients of the
realized curvature, $\kappa$, are then 
% $ \alpha_k = \alpha_k^{\infty}
%  \left(1-\left(1-\frac{\G\mu_k}{2\pi i}\right)^{-1}\right).$
%
%
\begin{equation}\label{visc_curv}
  \alpha_k = \alpha_k^{\infty}
  \left(1-\left(1-\frac{\G\mu_k}{2\pi i}\right)^{-1}\right).
\end{equation}
From this solution we can see that as the rod is stiffened
($\G\rightarrow\infty$), the resultant curvature tends to the
prescribed curvature, $\alpha_k\rightarrow\alpha_k^{\infty}.$ We also
see that for softer rods, i.e.\ smaller values of the body response time $\G$, the amplitude
of the curvature decreases and there is a phase lag relative to the
prescribed shape. 

\textcolor{review_color}{As mentioned above, we use intrinsic coordinates and curvature deviations, to allow us to consider large prescribed curvatures. However we note that equation \eqref{curveq} lacks terms coming from geometric nonlinearities and inextensibility that may not be small when  
the prescribed curvature is large \cite{PhysRevLett.75.1094,camalet2000generic}. In sections \ref{flex_compare} and \ref{ss_compare}
 we compare our simulations to theoretical analysis using \eqref{curveq}, and in Appendix \ref{App:largeamp} we show that the influence of the additional terms is in fact small for the amplitudes we consider.}

%%%%%%%%%%%%%%%%%%%%%%%%%%%%%%%%%%%%%%%%%%%%%%%%%%%%%%%%%%%%%%%%%%%%%%%%%%%

\subsection{Linear theory: viscoelastic fluids}\label{flex_ve}
We can modify the linear theory for elastic rods to include fluid
elasticity. This is similar to what was done in
\cite{fulford1998swimming,fu2008beating}.  \textcolor{review_color}{In \cite{fulford1998swimming} 
modifications to linear rod theory to include linear viscoelastic fluid effects were presented, and the authors concluded that while fluid elasticity does not change swimming speed, it reduces total work and thus can boost efficiency. However, it was pointed out in \cite{LAUGA:2007}
that it is essential to use a nonlinear elasticity model in these types of calculations because the swimming speed itself is second order in amplitude, where the nonlinear effects are relevant.  We note that with these higher order terms \cite{LAUGA:2007} shows that swimming speed is always hindered by fluid elasticity for the case considered \--- infinite length low amplitude swimmer with sinusoidal undulations.
In \cite{fu2008beating} the authors analyzed shape changes 
induced by fluid elasticity in a linearly elastic fluid.
Unlike swimming speed, shape changes due to fluid elasticity come in to the asymptotic expansion at first order in amplitude, and hence it is 
   reasonable to use a linearly elastic fluid to look at shape changes. \cite{fu2008beating} did not make conclusions about how these
shape changes affect swimming speed.  Here we perform a similar analysis as in \cite{fu2008beating}, but by applying the analysis to deviations in curvature we are able to study shape changes in low \textit{and high} amplitude finite length flexible rods.  In Section \ref{sec:ana_ss} we discuss how these shape changes affect swimming speed. }

As in equation
\eqref{beameq} we can write a force balance relation between the force
on a fluid and from the beam as
  \[\ff-\kbdim(y_{ss}-\kappa_0)_{ss}=0,\]
where the $\ff$ represents the normal force on the rod from the
viscoelastic fluid. If we define the fluid force to be based on the
total deviatoric stress $\boldsymbol{\tau}=\dgamma+\taup$ then (upon
linearization) using equation \eqref{stresseq}: 
%\footnote{Note that given the form of the system in
%  equations \eqref{mombaleq}\--\eqref{stresseq} we have assumed a total
%  viscosity of $1+\xi.$
%}
 \begin{equation}\label{forcebal}
   \De\;\dot{f}_{\textrm{fluid}}+\ff =
   (1+\xi)\fv+\De\;\dot{f}_{\textrm{vis}},
 \end{equation} 
 where $\fv$ is the viscous drag force.
Note that given the form of the system in equations
\eqref{mombaleq}\--\eqref{stresseq}, we have assumed a total viscosity
of $1+\xi$.
The swimmer motion is time-periodic so we take the Fourier transform
in time of equation \eqref{forcebal} to solve for viscoelastic
modifications to the fluid drag.  This yields, $ \hat{f}_{\textrm{fluid}}=
  \left(\frac{1+\xi+2\pi i\De}{1+2\pi i \De}\right)\hat{f}_{\textrm{vis}}.$
%\[
%  \hat{f}_{\textrm{fluid}}=
%  \left(\frac{1+\xi+2\pi i\De}{1+2\pi i \De}\right)\hat{f}_{\textrm{vis}}.
%\]

%Note that often in the literature the polymer to total viscosity is
%defined as $\beta=\frac{\eta_p}{\eta_s+\eta_p},$ which is related to
%$\xi$ by $\beta=\frac{1}{1+\xi}.$ 
As in the viscous theory, we can solve for modifications to the
curvature from body stiffness and use the modifications to the
fluid drag to account for the fluid elasticity:
\begin{equation}\label{veshapechanges}
  \alpha_k = \alpha_k^{\infty}
             \left(1-\left(1-\frac{\G\mu_k}{\textcolor{review_color}{\zetave}2\pi i}\right)^{-1}\right),\textrm{ with }\;
 \textcolor{review_color}{ \zetave}=\frac{1+\xi +2\pi i\De }{1+ 2\pi i\De }.
\end{equation}
The
coefficients in equation \eqref{veshapechanges} give an analytical
expression for the modifications to the rod shapes relative to the
prescribed shapes as fluid and body elasticity are varied.

%\textcolor{review_color}{, and here it can be seen
%that the effect of fluid elasticity affects the shapes (given by the $\alpha_k$) at all
%orders.}

%, note that
%shape changes are first order in the prescribed amplitude. 
%(DO WE WANT TO NOTE THAT, WE DO	NOT COME
%BACK TO IT AGAIN, SHOULD WE?)

%%%%%%%%%%%%%%%%%%%%%%%%%%%%%%%%%%%%%%%%%%%%%%%%%%%%%%%%%%%%%%%%%%%%%%%%%%%

\subsection{Elastic shape changes: theory and numerical comparison}
 \label{flex_compare}
The analysis in the previous sections made use of resistive force theory which
relates the drag force and velocity on a long thin cylindrical object.  More generally, for small amplitude
 the vertical displacement satisfies \begin{equation}\label{linbeam}y_t = M F y,\end{equation} where $M$ is the mobility operator and $F$ is the linearized
bending force operator.  Resistive force theory makes the approximation $M\approx\frac{1}{\zeta_{\perp}}.$ Our 
analysis of shapes (equation \eqref{veshapechanges}) contains the quantity $\G\mu_k,$ where $\G$ depends
 on the drag coefficient, $\zeta_{\perp},$ (equation \eqref{Gdef}). To use the more general linear theory in our 
 analysis one can identify $\mu_k/\zeta_{\perp}=\mu_k^{MF}$, where $\mu_k^{MF}$
 denotes the $k^{th}$ eigenvalue of the operator $MF.$ 
We relate $\G\mu_k$ to the dimensionless bending stiffness, $\kb$ used in our simulations, through 
\[\G\mu_k=\frac{T\kbdim}{L^4\zeta_{\perp}}\mu_k=\left(\frac{T\kbdim}{L^4}\right)\left(\frac{\mu_k}{\zeta_{\perp}}\right)=\kb\mu_k^{MF}.\]

\textcolor{review_color2}{To compare the linear analysis with our two-dimensional simulations we numerically approximate equation \eqref{linbeam}.
For small deviations to the vertical displacement, $M$ is the integral operator which is the convolution of the vertical force with the 
fundamental solution to Stokes equations. We approximate $M$ using the method of regularized Stokeslets
\cite{cortez2001method}, which is a numerical method based on a regularized Greens function for the Stokes equations.
%To compare with our simulations we compute $\mu_k^{MF}$ from $MF$ by approximating the mobility matrix, $M,$
%using the method of regularized Stokeslets
%\cite{cortez2001method}, which is a numerical method based on a regularized Greens function for the Stokes equations.
 We can also 
numerically approximate $\mu^{F}_{k},$ the $k^{th}$ eigenvalue of the bending force operator $F,$ using a second-order finite difference, and we find that with point spacing $\triangle s = 0.002,$
the eigenvalues of $F$ are within 1\% of the
eigenvalues of the continuous operator.  We give the eigenvalues for the first four nontrivial modes in table \ref{tab:evals}.  Note that to compute $\mu_k^{MF}$ we assume viscosity one.
Also in  table \ref{tab:evals}, we give the first four (mode dependent) drag coefficients computed as $\zeta_{k} = \mu^{F}_{k}/\mu_{k}^{MF}$.}

%With these eigenvalues we can  also 
%compute a (mode dependent) drag coefficient $\zeta_{k} = \mu^{F}_{k}/\mu_{k}^{MF},$ which are also given in table \ref{tab:evals}.}

% For the first four nontrivial
%modes, $k=1-4,$ the corresponding eigenvalues are $\mu_{k}^{F}$=
%\{$-$4.97e+02, $-$3.77e+03, $-$1.45e+04, $-$3.96e+04\}, and for viscosity one,
%$\mu_{k}^{MF}$=\{$-$1.53e+01, $-$8.79e+01, $-$2.65e+02 , $-$5.93e+02\}.     Note that with these values we can also 
%compute a (mode dependent) drag coefficient $\zeta_{k} = \mu^{F}_{k}/\mu_{k}^{MF},$ and for the  first four nontrivial modes the drag coefficients are $\zeta_{k}$=\{32.55, 42.94, 54.69,
%66.84\}.  }

\begin{table}
\begin{center}
\begin{tabular}{c c c c}
Mode ($k$) & $\mu_{k}^{F}$ & $\mu_{k}^{MF} $ &$ \zeta_{k}$  \\\hline\hline
1 & $-4.97\times10^2$& $-1.53\times 10^1$& 32.55\\\hline
2 & $-3.77\times10^3$ & $-8.79\times 10^1$ &42.94 \\\hline
3 & $-1.45\times 10^4$& $-2.65\times 10^2$ & 54.69\\\hline
4 & $-3.96\times 10^4$ & $-5.93\times 10^2$ & 66.84\\\hline
\end{tabular}
\end{center}\caption{Eigenvalues of the discretized operators $F$ and $MF$ using $\triangle s=0.002,$ and the effective drag coefficient $\zeta_{k} = \mu^{F}_{k}/\mu_{k}^{MF}$ for the first four non-trivial modes.}
\label{tab:evals}
\end{table}

\begin{figure}
\centering
\includegraphics[width=\textwidth]{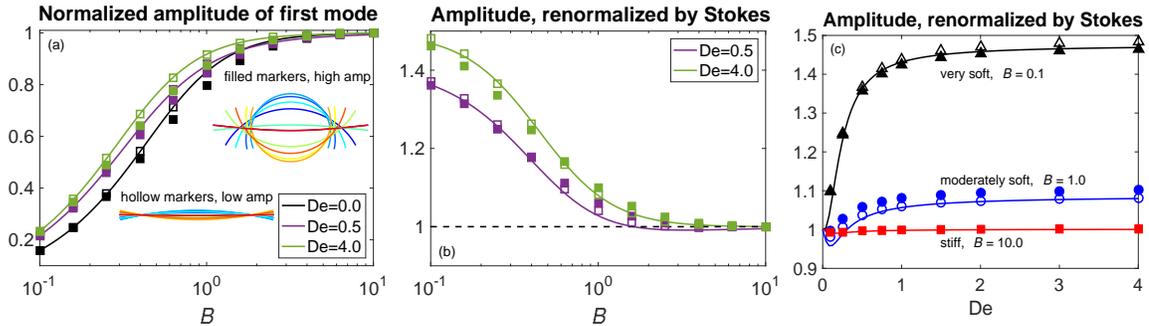}
\caption{(a) Normalized amplitude of the first mode,  $|\alpha_1|/|\alpha^{\infty}_1|,$ from
equation \eqref{veshapechanges}, for the flexor at low $A=0.5$ and high $A=4.0$ amplitude over a range
of bending stiffness $\kbdim.$ Linear theory is shown in solid lines and simulation data are indicated by markers. 
(b)-(c) Data from (a) renormalized by Newtonian ($\De=0$) data, as a function of both $\kbdim$ and $\De.$}\label{fig_shapes_IB}
\end{figure}

In order to compare the predicted shape changes given by equation \eqref{veshapechanges} with our numerical simulations we prescribe a curvature of the
form 
\begin{equation}\label{flex_curv}
  \kappa_0(s,t)=A\sin(2\pi t), \end{equation}
in our model equations \eqref{mombaleq}--\eqref{eq_curve}. 
The prescribed standing wave of constant curvature corresponds to a
motion through circular arcs with peak curvature $A$.
By symmetry, this motion does not result in any horizontal translation of the body.
We refer to these non-translating ``swimmers" as {\em flexors}. We
consider both low and high amplitude curvatures, $A=0.5$ and $A=4.0.$
The shapes are shown inset in figure \ref{fig_shapes_IB}(a).

%\begin{figure}
%\centering
%\includegraphics[width=\textwidth]{figure4.eps}
%\caption{(a) Normalized amplitude, theoretical (lines) and simulations (markers).
%(b)-(c) Data renormalized by Stokes.}\label{fig_shapes_IB}
%\end{figure} 

In figure \ref{fig_shapes_IB} we plot the theoretical predictions from
equation \eqref{veshapechanges} (solid lines) along with values
computed from numerical simulations; low amplitude ($A=0.5$) are
indicated by hollow markers, and high amplitude ($A=4.0$) are
indicated with filled markers.  In figure \ref{fig_shapes_IB}(a) we
plot the normalized amplitude of the first mode
($|\alpha_1|/|\alpha_1^{\infty}|$) to see how the amplitude deviates
from the prescribed amplitude as a function of bending stiffness $\kb$. We see that generically the amplitude of the
flexor decreases as the flexor is softened for fixed $\De$.  For
sufficiently soft flexors ($\kb\lesssim 1$) viscoelasticity increases the
amplitude monotonically with $\De$, but for stiffer swimmers the
amplitude changes nonmonotonically with fluid elasticity.

In figures \ref{fig_shapes_IB}(b) and (c) we renormalize the data by the
amplitude in a viscous fluid to see the effects of
viscoelasticity more clearly.  Again we see that fluid elasticity can
increase the amplitude significantly for a soft flexor, but that effect
is lost as the flexor is stiffened.  When we plot the amplitude as a
function of $\De$ for the {\em very soft, moderately soft} and {\em
  stiff} cases we see again that three qualitatively different regimes emerge. 
  For {\em very soft} flexors the amplitude is
monotonically increased by elasticity, for {\em moderately soft}
flexors the response is non-monotonic, and can decrease or increase the amplitude, and for
{\em stiff} flexors there is little change in the amplitude due to fluid
elasticity.  It is notable that the linear theory does such a good job
predicting shape changes for low and high amplitude and for low and
high Deborah number.  \textcolor{review_color}{In Appendix \ref{App:largeamp} we derive the theory for both the limit of small amplitude
and the limit of high stiffness.  We see in figures \ref{fig_shapes_IB} (a) and (b) that the largest differences are for moderate stiffnesses at high amplitude.} We note that we are showing results only for the first mode.  For
higher modes the trends are similar but the transition from stiff to
soft behavior occurs at lower values of $\kb$ because the eigenvalues
$\mu_{k}$ increase with $k$. 
%
%{\color{blue} WHY SAY THE NEXT SENTENCE: The exact details of these curves also
%depends on the viscosity ratio which appears in equation
%\eqref{veshapechanges}; results here are shown for $\xi=0.5.$}

%%%%%%%%%%%%%%%%%%%%%%%%%%%%%%%%%%%%%%%%%%%%%%%%%%%%%%%%%%%%%%%%%%%%%%%%%%%
%
% Section 4: Swimming Speed
%
%%%%%%%%%%%%%%%%%%%%%%%%%%%%%%%%%%%%%%%%%%%%%%%%%%%%%%%%%%%%%%%%%%%%%%%%%%%

\section{Analysis of swimming speed}\label{sec:ana_ss}

In a viscous fluid, increasing the stroke amplitude will increase the swimming speed, and we can infer from section \ref{Sec_shape} that soft swimmers in a viscoelastic fluid sometimes obtain an amplitude boost over the corresponding swimmer in a Newtonian fluid. However when comparing swimmers in a viscoelastic fluid to those in a viscous fluid, even with an amplitude boost the viscoelastic swimmer may not swim faster than the viscous swimmer due to additional fluid elastic forces that the swimmer will encounter.   Thus the effect of elasticity-induced shape changes is difficult to decouple from the overall effect of fluid elasticity. Analytical expressions for swimming speed can be obtained in certain limits, or for specialized swimmers, but even in these cases we see that the effect of fluid elasticity depends on many factors. For example infinite-length small amplitude undulatory swimmers show that a slow-down is generically expected for stiff swimmers in a viscoelastic fluid \cite{LAUGA:2007},
 but allowing for body flexibility, shape changes can lead to speed boosts \cite{riley2014enhanced}.   
 
  In regimes that are more challenging for analysis such as the large amplitude, finite length swimmers considered here, it is more difficult to attribute speed boosts or slow-downs to specific swimmer attributes.  
For large amplitude finite-length undulatory swimmers, it was conjectured \cite{teran2010viscoelastic} that speed boosts were related to large tail stresses,  and in \cite{Thomases2014d}  stroke asymmetries were correlated with both slow-downs and speed-ups. Here we will compute a \textit{stroke-induced swimming speed} that isolates the effect of fluid elasticity on shape changes, and how those shape changes affect swimming speed in a Newtonian fluid.  We then compare that analysis with the full nonlinear numerical simulations where the effect of shape changes is coupled with the fluid elasticity.

\subsection{Swimming speed: two-mode swimmer}\label{sec:2modeswim}
\begin{figure}
\centering
\includegraphics[width=.85\textwidth]{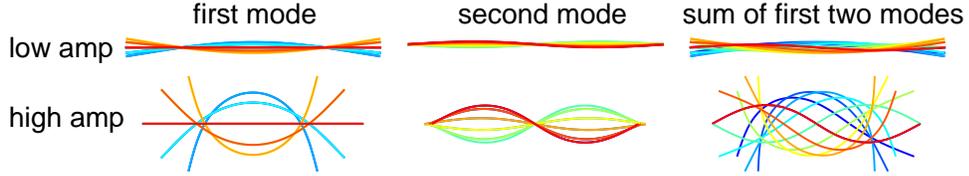}
\caption{Shapes for first, second, and sum of first two modes for ``two-mode" swimmer defined by equation \eqref{twomode_eq}. Snapshots of one period for low and high amplitude strokes.}\label{2mode_shapes}
\end{figure}

To keep
the analysis simple we define a gait whose swimming speed in a viscous fluid we can compute analytically.  We define a
``two-mode swimmer" given by the curvature: 
%\begin{equation}
%  \kappa(s,t)=\alpha_1\cos(2\pi t/T+\phi_1)\Psi_1(s)
%             +\alpha_2\cos(2\pi t/T+\phi_2)\Psi_2(s),
%\label{twomode_eq}
%\end{equation} 
\begin{equation}
  \kappa(s,t)=A_1\cos(2\pi t/T+\phi_1)\Psi_1(s)
             +A_2\cos(2\pi t/T+\phi_2)\Psi_2(s),
\label{twomode_eq}
\end{equation} 
where the $\Psi_i(s)$ for $i=1,2,$ are the first and second bending modes.  The
modulation of a single mode results in a standing wave and will not
translate in a Newtonian fluid.  We use a sum of the first two modes
with a phase difference to generate a nonreciprocal motion. Shapes of
the first, second, and sum of the first and second modes are plotted
in figure \ref{2mode_shapes} for both low and high amplitudes.
 
Using resistive force theory one can derive the (time-averaged)
swimming speed for a given, small amplitude, motion:
\begin{equation}
 \label{ss_eq}
 \ip{U}=\left(\frac{\zeta_{\perp}}{\zeta_{\parallel}}-1\right)\frac{1}{LT}\int_0^T\int_0^Ly_sy_t\;dsdt,
\end{equation} 
where $U$ is the swimming speed, $y(s,t)$ is the vertical displacement of
the swimmer, and $\zeta_{\perp}$ and $\zeta_{\parallel}$ are the
perpendicular and parallel drag coefficients, respectively,
\cite{wiggins1998flexive,lauga2007floppy}.

For small
amplitudes, the shape of the swimmer (up to translation and rotation)
is given by integrating equation \eqref{twomode_eq} twice in space to
compute the swimming speed via equation \eqref{ss_eq}. 
 The swimming
speed (in a viscous fluid) for the two-mode swimmer is proportional to the
product of the amplitudes and the sine of the phase difference: \begin{equation}\label{stroke_speed}\ip{U}\propto  A_1A_2\sin(\phi_2-\phi_1).\end{equation}
%
%Using the expressions that
%give the phase and amplitude dependence on $\De$ and $\G$
%from equation \eqref{veshapechanges}, equation \eqref{stroke_speed} then gives the swimming speed in a Newtonian
 %fluid that depends on the shape changes due to fluid elasticity.  
 \textcolor{review_color2}{With this expression and the theoretical prediction for shape changes, we 
define a stroke-induced swimming speed which is the swimming speed in a Newtonian fluid that
depends on the shape changes due to fluid elasticity and body flexibility. Specifically, for a given $\De$ and $\G,$ we compute $\alpha_k$ from equation \eqref{veshapechanges} ($A_j=|\alpha_j|,$ and $\phi_j=\arg(\alpha_j)$) and the stroke-induced swimming speed from equation \eqref{stroke_speed}.  
We parametrize the shape changes due to changes in $\De$ using a parameter we call the stroke-Deborah number \cite{Thomases2014d},  $\strDe.$ In other words, $\strDe$ represents the value of $\De$ used in equation \eqref{veshapechanges} to compute the stroke-induced swimming speed via equation \eqref{stroke_speed}.}
Our analytical expression for the shape changes is based on a linearly elastic fluid, but because the nonlinear elastic effects and swimming speed are both second order in amplitude, we do not expect the stroke-induced swimming speed to capture the true viscoelastic swimming speed.
 An analytical expression for the swimming speed in a nonlinear viscoelastic fluid, as was computed in \cite{LAUGA:2007}, is not tractable in the finite length case, because translational invariance, which facilitates the calculation for infinite length swimmers, is lost.

We plot the stroke-induced swimming speed for the two-mode swimmer over a range of $\strDe,$  and as with the flexor, we see the emergence of three regimes dependent on the body stiffness, see figure \ref{twomode_snss} (a).  Shape changes boost the stroke-induced swimming speed if the swimmer is \textit{very soft}, a smaller boost is obtained for the \textit{moderately soft} swimmer, and additionally there is a non-monotonic response to increasing elasticity including a regime where shape changes slow down the swimmer, and finally if the swimmer is \textit{stiff} there is a negligible effect. 

\subsection{Swimming speed: theory and numerical comparison}\label{ss_compare}

\begin{figure}
\centering
\includegraphics[width=.75\textwidth]{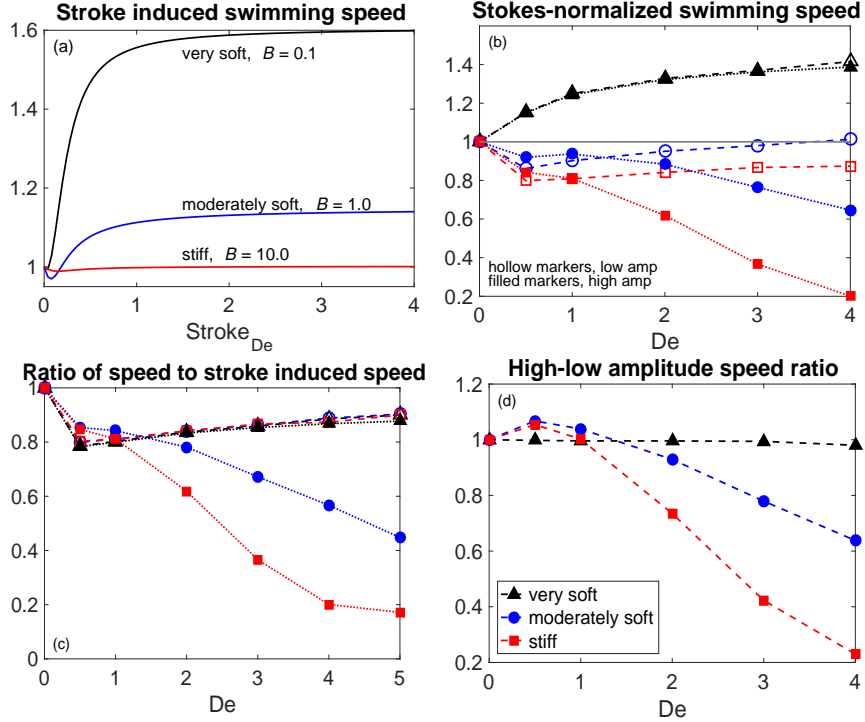}
\caption{(a) Theoretically predicted stroke-induced swimming speed (computed using equation \eqref{stroke_speed}) normalized by the Newtonian stroke-induced swimming speed. (b) Stokes-normalized swimming speed in simulations with the two-mode swimmer. (c) Ratio of speed to stroke-induced speed for low and high amplitude strokes. (d) Ratio of high to low amplitude swimming speed.  (Dashed lines are for  graphical interpretation).}\label{twomode_snss}
\end{figure} 

We simulate a two-mode swimmer of both low and high amplitude by
prescribing a curvature of the form given in equation \eqref{twomode_eq} 
with $A_1=0.8A$, $A_2=0.6A$, $\phi_2-\phi_1=\pi/2$, for
$A=0.5$ (low), and $4.0$ (high). These values come from projections
of the stroke used to generate figure \ref{fig_vg}.  The Stokes-normalized swimming speeds for a \textit{very soft, moderately soft}, and \textit{stiff} swimmer at both low (hollow markers) and high (filled markers) amplitude are plotted in figure \ref{twomode_snss} (b).
We note that the three regimes seen in figure \ref{twomode_snss} (a) still emerge from these simulations, but \textcolor{review_color}{for this two-mode swimmer the simulation swimming speeds are always  slower than the stroke-induced swimming speeds}. The ratio of swimming speed to stroke-induced swimming speed is shown in figure \ref{twomode_snss} (c). This quantity can be interpreted as the effect of fluid elasticity that is \textit{not} related to shape changes. It is notable that these curves collapse onto a single curve for the low amplitude swimmers at all stiffnesses as well as the high-amplitude swimmer in the very soft regime.  
This additional elastic fluid effect on swimming speed is likely to be highly stroke dependent. %A similar idea was introduced in \cite{LAUGA:2007},
%and further analysed in \cite{riley2014enhanced} for infinite-length low amplitude swimmers. Our work predicts a much different dependence on $\De$
%than what was seen in \cite{LAUGA:2007,riley2014enhanced}.  

\begin{figure}
\centering
\includegraphics[width=.95\textwidth]{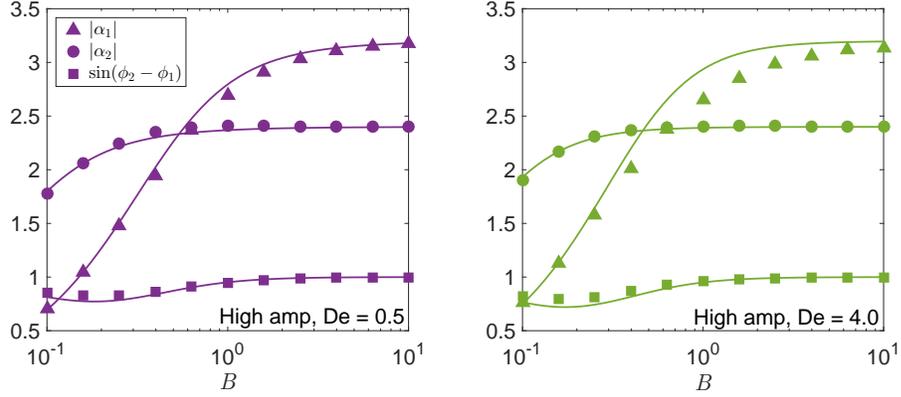}
\caption{(a) Amplitudes $|\alpha_1|, |\alpha_2|,$ and $\sin(\phi_2-\phi_1)$, for the two-mode swimmer, equation \eqref{twomode_eq}).  Linear theory is shown in solid lines (dependence on $\kbdim,$ and $\De$ coming from equation \eqref{veshapechanges}) and simulation data are indicated by markers for high amplitude $A=4.0,$ and $\De = 0.5$ (a), $\De = 4.0$ (b).}\label{IBth_2mode}
\end{figure} 

The additional effects of fluid elasticity are fundamentally different for the large amplitude, large $\De$ regime.  In figure \ref{twomode_snss} (d) we plot the ratio of swimming speeds for the high-to-low amplitude strokes, and see that for  $\De> 1$ (for sufficiently stiff swimmers) a significant difference in swimming speed arises.  This difference is not related to shape changes because, like the flexor, the elasticity-induced shape changes predicted by the theory for the two-mode swimmer agree very well with the simulation results, for all $\De,$ at low and high amplitudes; see figure \ref{IBth_2mode}. 
At low amplitude (not shown) the relative error between theoretical and numerical predictions is less than 1$\%,$ and for high amplitude the error at low $\De$ is at most 5$\%$ and at high $\De$ the error is at most 9$\%$.  These results indicate that the theoretically predicted shape changes and their isolated effects on swimming speed can be well approximated by the analytical results for the amplitudes simulated and the range of $\De$ considered. A mechanistic understanding is lacking to explain what causes the dramatic slow-downs of swimmers in the high amplitude, high $\De$ regime.

\begin{figure}
\centering
\includegraphics[width=.85\textwidth]{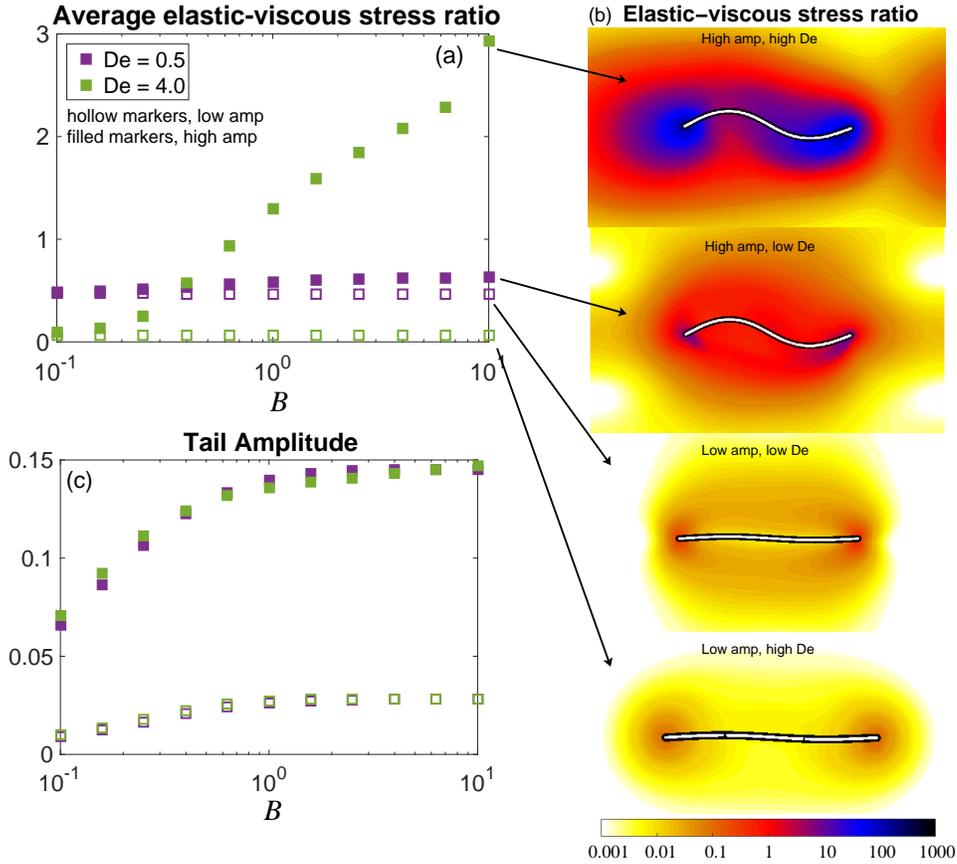}
\caption{(a) Average elastic-viscous stress ratio in two-mode swimmer over a range of bending stiffness $\kbdim$
for both low and high amplitude strokes and low and high $\De.$  (b) Color field of the magnitude of the elastic-viscous stress ratio around stiff swimmers ($\kbdim=10$) at low and high $\De$ and low and high amplitudes. (c) Tail amplitudes over a range of bending stiffness $\kbdim$, measured as the maximum displacement at the tail over a period, for both low and high amplitude strokes and low and high $\De.$}\label{fig_stress}
\end{figure} 

We conjecture that the slow-downs in the high amplitude, high $\De$ regime must be attributed in part to
the large localized stresses that accumulate near the body  \cite{teran2010viscoelastic,Thomases2014d}. To explore this conjecture, in figure \ref{fig_stress} (a) we plot the average elastic to viscous stress ratio\textcolor{review_color2}{, computed as the time average over one period of $\displaystyle\|\taup\|/\|\nabla\u\|$ where $\|\cdot\|$ is the Frobenius norm,}
over a range of body stiffness $\kb$ for
    both low (hollow markers) and high amplitude (filled markers)
strokes for $\De = 0.5, 4.0.$  There is a notable transition in the stress ratio in the high $\De$, high amplitude
swimmer as the body is stiffened, while this stress ratio is flat for both low amplitude and low $\De$ swimmers.  \textit{Stiff} swimmer shapes along with the elastic-viscous stress ratio are plotted on a log-scale in figure \ref{fig_stress} (b). The low amplitude strokes are surrounded by elastic stresses that are at least two orders of magnitude smaller than the high amplitude strokes, but even at large amplitude the low $\De$ swimmer still has relatively low stress near the body. 
 Lastly, we plot the tail amplitude, as one measure of the swimmer stroke, in figure \ref{fig_stress} (c). We see that for sufficiently soft swimmers
the ``high amplitude" stroke has a lower amplitude, which explains why in figure \ref{twomode_snss} the \textit{very soft} high amplitude swimmer behaves like the low amplitude swimmers. For the high amplitude strokes it is only in the large amplitude \textit{and} large $\De$ regime where significant stress accumulates near the swimmer.

%%%%%%%%%%%%%%%%%%%%%%%%%%%%%%%%%%%%%%%%%%%%%%%%%%%%%%%%%%%%%%%%%%%%%%%%%%%
%
% Section 5
%
%%%%%%%%%%%%%%%%%%%%%%%%%%%%%%%%%%%%%%%%%%%%%%%%%%%%%%%%%%%%%%%%%%%%%%%%%%%

\section{Conclusions}
In \cite{Thomases2014d} we showed that stroke related speed-ups depend on
body stiffness, and the analysis from this paper shows explicitly how
the stroke changes depend on body stiffness and fluid elasticity through 
two dimensionless ``relaxation times": the fluid relaxation time, $\De$ and the body
relaxation time, $\G^{-1}.$  When we look at apparently contradictory results from the 
literature, we see that calculating $\G$ will determine which regime the swimmer falls into.

In
\cite{espinosa2013fluid} the Sperm number is reported to be between
$0.5-2.5,$ but even with the awareness that these are soft swimmers
the authors ``conjecture that the effect [due to shape changes] is not
significant". We use their reported
parameters\footnote{$L=25$mm, cross-sectional radius $a=62.5 \mu$m,
  Young's modulus $E=80$ GPa, viscosity $\mu=2.7$ Pa-s. For moment of
  inertia $I=\pi a^4/4,$ we get $\kbdim=E I=9.6\times 10^{-7},$
  $\zeta=4\pi\mu/\ln(L/a),$ and thus $\G = \frac{\kbdim T}{\zeta
    L^4}=\frac{9.6\times 10^{-7}\ln(400)}{4\pi\cdot
    2.7(25\times10^{-3})^4}\approx 0.43.$} and a characteristic
frequency of $1$s$^{-1},$ and find $\G\approx 0.43.$ \textcolor{review_color}{We
cannot directly conclude that this value lies in the \textit{very soft} regime ($\kb\approx 0.1$) due
to differences between 2D and 3D as well as the way that the micro-swimmer is driven (it is a 
flexible tail with a magnetically driven head). In Appendix \ref{App:enddriven} we repeat the calculation from section \ref{sec:2modeswim} 
to compute an equivalent stroke-induced swimming speed for a flexible filament driven at one end.  This calculation shows that speed boosts still arise for sufficiently soft swimmers, despite the difference in driving mechanism. The most significant boost in speed from viscoelastic shape changes occurs for $\G\approx 0.1,$ but for $\G\approx 0.43$ it would be reasonable
to conclude that the significant speed-ups observed in the experiment are related to shape changes. }

In \cite{salazar2016numerical} the parameter reported for
what they consider to be a soft swimmer is $\kb=2$. However
their swimmer length is $L=0.6$ mm (with characteristic length 1 mm) hence an equivalent dimensionless body
response time $\G$ must be multiplied by $L^{-4}\approx 7.7$. This
pushes their ``soft" simulations into the \textit{stiff} regime where
there are no speed-ups from shape changes, also agreeing with their results. Furthermore, in \cite{salazar2016numerical}
it is conjectured that stress diffusion, used to regularize the simulations in \cite{teran2010viscoelastic,Thomases2014d},
is the source of the speed-ups, but the speed-ups we see are theoretically
predicted, and realized in our simulations, even in the low amplitude regime where
no regularization is necessary.
 
In our analysis we quantify the effect of body and fluid elasticity-induced shape changes on swimming speed. 
We see that the shape change analysis holds for \textcolor{review_color}{the amplitudes simulated and the range of $\De$ considered}, and  \textcolor{review_color}{in this case}  we
see an additional elastic slow-down that is reminiscent of the type of slow-down predicted
by asymptotic analysis of infinite-length small amplitude undulatory swimmers \cite{LAUGA:2007}.  It may be tractable to apply
asymptotic analysis \cite{riley2014enhanced,elfring2016effect} to determine
the form of the elastic slow-down for low amplitude finite-length swimmers.  
A fundamentally different regime arises for large amplitude swimmers in highly elastic 
fluids. A different approach is needed to understand 
the mechanisms that cause large localized stresses and their effect on swimming.

%\begin{acknowledgments}
\vspace{.1in}
The authors would like to thank Henry Fu and Roberto Zenit for interesting discussions and suggestions on this work, and Michael Shelley for suggesting the term ``flexors".  \textcolor{review_color}{The authors would also like to thank the anonymous referees for suggesting useful modifications to our original manuscript.} The work of RDG was partially supported by NSF grants
DMS-1160438 and DMS-1226386.%
%\end{acknowledgments}

\appendix

\section{Derivation of PDE for rod dynamics}\label{App:largeamp}

% notation I created for this appendix
%
\newcommand{\nv}{\hat{\bm{n}}}          % normal vector
\newcommand{\tv}{\hat{\bm{\tau}}}       % tangential vector
\newcommand{\Fn}{F_{\perp}}
\newcommand{\Ft}{F_{\parallel}}
\newcommand{\zn}{\zeta_{\perp}}
\newcommand{\zt}{\zeta_{\parallel}}

In this appendix we give the derivation for the equation of motion for
a thin filament in a viscous fluid which includes terms arising from
inextensibility and geometric nonlinearties. For more details on
similar calculations see for example
\cite{PhysRevLett.75.1094,camalet2000generic}.

% First, a description of the variables for shape
%
\subsection{Geometric Relations}
Consider an inextensible thin rod whose centerline position is given
by $\X(s,t)$, where $s$ is arclength coordinate. We suppose that the
deformation of the rod is planar.  Let $\tv$ and $\nv$ be the tangent
and normal vectors, $\psi$ be the tangent angle, and $\kappa$ be the
curvature. We have the following relationships between these quantities:
\begin{equation}
  \psi_{s} = \kappa, \quad
  \tv_{s} = \kappa \nv, \quad
  \nv_{s} = -\kappa\tv, \quad
  \tv_{t} = \nv\psi_{t}, \quad
  \nv_{t} = -\tv\psi_{t}.
\end{equation}
%\begin{gather}
%   \psi_{s} = \kappa \\
%   \tv_{s} = \kappa \nv \\
%   \nv_{s} = -\kappa\tv \\
%   \tv_{t} = \nv\psi_{t} \\
%   \nv_{t} = -\tv\psi_{t}
%\end{gather}

% derive equations of motion give forces
%
\subsection{Equation of Motion}
Let $\F(s,t)=\Fn\nv + \Ft\tv$ be the force density along the
rod. Using resistive force theory, the motion of the rod is given by
\begin{equation}
  \X_{t} = \frac{1}{\zn}\Fn \nv + \frac{1}{\zt}\Ft\tv,
\end{equation}
where $\zn$ and $\zt$ are the normal and tangential drag coefficients,
respectively. Taking the derivative of this equation with respect to
arclength gives
\begin{equation}
  \partial_{t}\X_{s} =   \left(\frac{1}{\zn}\partial_{s}\Fn 
                      + \frac{1}{\zt}\Ft \kappa\right) \nv
                      + \left(\frac{1}{\zt}\partial_{s}\Ft
                      - \frac{1}{\zn}\Fn \kappa\right)\tv.
\end{equation}
Because $\X_{s}=\tv$, the left side of the above equation can be
expressed as $\partial_{t}\X_{s}=\tv_{t} = \nv\psi_{t}$, and thus  
the normal terms give $\psi_{t}$ and the
tangential terms must be zero:
\begin{gather}
  \psi_{t} = \frac{1}{\zn}\partial_{s}\Fn + \frac{1}{\zt}\Ft \kappa, \label{psit:eq} \\
  \frac{1}{\zt}\partial_{s}\Ft - \frac{1}{\zn}\Fn \kappa =0. \label{inext:eq}
\end{gather}
Equation \eqref{inext:eq} represents a constraint on the forces that
must be satisfied to maintain inextensibility. The evolution equation
for the curvature is obtained by differentiating equation
\eqref{psit:eq} with respect to arclength to obtain
\begin{equation}
  \kappa_{t} = \frac{1}{\zn}\partial_{ss}\Fn + \frac{1}{\zt}\partial_{s}\left(\Ft \kappa\right).
  \label{kappat:eq}
\end{equation}

% Now derive forces for our constitutive low
%
\subsection{Expression for Elastic Forces}
The elastic forces are obtained from the variation of an elastic
energy functional.  The total elastic energy is the sum of a bending
term from equation \eqref{bending_energy:eq} with an energy associated
with inextensibility:
\begin{equation}
  \mathcal{E} = \int_{0}^{L} \frac{\kbdim}{2}(\kappa-\kappa_{0})^2 + \frac{\Lambda}{2}\X_{s}\cdot\X_{s} \, ds,
\end{equation}
where $\Lambda$ is a tension used to enforce inextensibility.  The
force comes from the variation of the energy
\begin{equation}
  \F = -\frac{\delta\mathcal{E}}{\delta\X}.
\end{equation}
Using the natural free boundary conditions
\begin{equation}
  \kappa = \kappa_{0}, \quad
  \kappa_{s} = \partial_{s}\kappa_{0}, \quad
  \Lambda    = 0,
\end{equation}
%\begin{gather}
%  \kappa = \kappa_{0} \\
%  \kappa_{s} = \partial_{s}\kappa_{0} \\
%  \Lambda    = 0,
%\end{gather}
the force is
\begin{equation}
 \F  = \frac{\partial}{\partial s}\biggl( -\kbdim(\kappa-\kappa_{0})_{s}\nv
                              + \left(\Lambda + \kbdim\kappa(\kappa-\kappa_{0}\right) \tv   \biggr) 
\end{equation}
We define the total tension in the rod as
\begin{equation}
  T = \Lambda + \kbdim\kappa(\kappa-\kappa_{0}), 
\end{equation}
and the expression for the force is
\begin{align}
  \F  & = \frac{\partial}{\partial s}\biggl( -\kbdim(\kappa-\kappa_{0})_{s}\nv+ T\tv   \biggr) \\
      & =  \bigl(-\kbdim(\kappa-\kappa_{0})_{ss} + \kappa T \bigr)\nv + 
              \bigl(\kbdim\kappa(\kappa-\kappa_{0})_{s} + T_{s} \bigr)\tv
\end{align}
The normal and tangential force densities on the rod are thus
\begin{align}
  \Fn & = -\kbdim(\kappa-\kappa_{0})_{ss} + \kappa T \\
  \Ft & = \kbdim\kappa(\kappa-\kappa_{0})_{s} + T_{s}.
\end{align}
With these forces, the evolution equation for the curvature
\eqref{kappat:eq} and the inextensibility constraint which
determines the tension \eqref{inext:eq} are
\begin{gather}
  \kappa_{t} = \frac{1}{\zeta_{\bot}} \bigl(-\kbdim(\kappa-\kappa_{0})_{ssss} + (\kappa T)_{ss} \bigr) +
  \frac{1}{\zeta_{\parallel}}\bigl(\kbdim(\kappa^2(\kappa-\kappa_{0})_{s})_{s}
  + (\kappa T_{s})_{s} \bigr),
  \label{curve_evol:eq} \\
  \frac{1}{\zt}T_{ss} - \frac{1}{\zn}\kappa^{2} T
  +  \frac{1}{\zt}\bigl(\kbdim\kappa(\kappa-\kappa_{0})_{s}\bigr)_{s}
  + \frac{1}{\zn} \kbdim\kappa(\kappa-\kappa_{0})_{ss} =0.
  \label{tensin:eq}
\end{gather}
These equations together with the boundary conditions at the ends of
the rod
\begin{equation}
  \kappa = \kappa_{0}, \quad
  \kappa_{s} = \partial_{s}\kappa_{0}, \quad
  T    = 0,
\end{equation}
%\begin{gather}
%  \kappa = \kappa_{0}, \\
%  \kappa_{s} = \partial_{s}\kappa_{0}, \\
%  T    = 0,
%\end{gather}
determine the motion.

%%%%%%%%%%%%%%%%%%%%%%%%%%%%%%%%%%%%%%%%%%%%%%%%%%%%%%%%%%%%%%%%%%%%%%%%%%%

\subsection{Asymptotic Expansions}
We examine the leading order behavior in two different limits: (1)
small amplitude motion in which $\kappa\rightarrow 0$, and (2) high
stiffness, $\kbdim\rightarrow \infty$, in which
$\kappa\rightarrow\kappa_{0}$.  Note that for soft bodies with large
$\kappa_{0}$ the realized amplitude, $\kappa$, is in fact small. Thus
we expect the largest discrepancy between the asymptotic solutions at
large $\kappa_{0}$ at intermediate stiffness. In fact this is what we
see in figure \ref{fig_shapes_IB} for the flexor and figure
\ref{IBth_2mode} for the two-mode swimmer.

\subsubsection{Small Amplitude}
In the limit of small curvature, we see from equation
\eqref{tensin:eq} that the size of the tension is like the square of
the curvature. At leading order the tension is zero, and the equation
for the curvature is
\begin{equation}
  \kappa_{t} = -\frac{\kbdim}{\zeta_{\bot}} (\kappa-\kappa_{0})_{ssss}.
\end{equation}
Changing to curvature deviation, $c=\kappa-\kappa_{0}$, and
nondimensionalizing gives equation \eqref{curveq} analyzed in the main
text.

\subsubsection{Large Stiffness}
In the limit $\kbdim\rightarrow\infty$, $\kappa\rightarrow\kappa_{0}$. We introduce the variable
\begin{equation}
  c = \kappa - \kappa_{0}
\end{equation}
to denote the deviation from the prescribed curvature.  For increasing
stiffness, $c\rightarrow 0$. Changing variables from $\kappa$ to $c$
and linearizing about small $c$ gives
\begin{gather}
  c_{t} + \partial_{t}\kappa_{0} = \frac{1}{\zn} \bigl(-\kbdim c_{ssss} + (\kappa_{0} T)_{ss} \bigr) +
                       \frac{1}{\zt}\bigl(\kbdim(\kappa_{0}^2 c_{s})_{s} + (\kappa_{0} T_{s})_{s} \bigr) \label{kappa_stiff_lin:eq} \\
 \frac{1}{\zt}T_{ss} -\frac{1}{\zn} \kappa_{0}^2 T + 
       \frac{1}{\zt}\kbdim\left(\kappa_{0} c_{s}\right)_{s} +
         \frac{1}{\zn}\kbdim\kappa_{0}c_{ss} = 0,
 \label{T_stiff_linear:eq}
\end{gather}
and the boundary conditions are
\begin{equation}
  c = 0, \quad c_{s} =0, \quad T    = 0.
\end{equation}
%\begin{gather}
%  c = 0 \\
%  c_{s} =0  \\
%  T    = 0.
%\end{gather}
These equations contain many terms that are absent for small
curvatures. However, as demonstrated in the text by comparing with
numerical results, the low curvature equations appear to give a
reasonable approximation at the amplitudes tested. Below we show why the
two approximations are similar.

For simplicity we consider the flexor in which $\kappa_{0}$ is only a
function of time. With this simplification, equations
\eqref{kappa_stiff_lin:eq}\--\eqref{T_stiff_linear:eq} become
\begin{gather}
  c_{t} + \partial_{t}\kappa_{0} = -\frac{\kbdim}{\zn} c_{ssss} +
  \frac{\kbdim\kappa_{0}^2}{\zt} c_{ss}
  + 
\kappa_{0}\left( \frac{1}{\zn} +  \frac{1}{\zt}\right) T_{ss}
  \label{kappa_stiff_linear_flex:eq}
 \\
 \frac{1}{\zt}T_{ss} -\frac{1}{\zn} \kappa_{0}^2 T + 
 \kbdim \kappa_{0}\left(\frac{1}{\zt}+ \frac{1}{\zn}\right)c_{ss} = 0.
  \label{T_stiff_linear_flex:eq}
 \end{gather}
These equations can be solved by orthogonal function expansion. First
we eliminate the tension by solving the second equation. We express
$T$ as the series
\begin{equation}
  T = \sum_{m=1}^{\infty} \beta_{m}(t) \sin\left(m\pi s\right).
\end{equation}
We then write the function $T$ as
\begin{equation}
  T(s,t) = Q^{-1} \vec{\beta},
\end{equation}
where $\vec{\beta}$ represents the sequence of coefficients and
$Q^{-1}$ is the orthogonal operator which maps these coefficients to
$T$, i.e.\ $Q$ is like the Fourier transform operator. Using this
expansion, the operator applied to $T$ in equation
\eqref{T_stiff_linear_flex:eq} diagonalizes, and the solution is
\begin{equation}
  T = \kbdim \kappa_{0}\left(\frac{1}{\zt}+ \frac{1}{\zn}\right)
  Q^{-1}\left( \frac{1}{\zt}M^{2} + \frac{1}{\zn} \kappa_{0}^2\right)^{-1}
  Q c_{ss},
\end{equation}
where $M$ is a diagonal matrix with elements $m\pi$ on the
diagonal. After using this expression to eliminate $T$ in equation
\eqref{kappa_stiff_linear_flex:eq}, after some simplification, we get
the equation for the curvature deviation as
\begin{equation}
  c_{t} + \partial_{t}\kappa_{0} = -\frac{\kbdim}{\zn} c_{ssss} +
  \frac{\kbdim\kappa_{0}^2}{\zt}Q^{-1} D Q c_{ss},
\label{curv_dev_simp:eq}
\end{equation}
where $D$ is a diagonal matrix with elements on the diagonal
\begin{equation}
  D_{mm} = 1 - \frac{\left(1+ \frac{\zt}{\zn}\right)^{2}}{\left(1 + \frac{\zt}{\zn} \frac{\kappa_{0}^2}{m^{2}\pi^{2}}\right)}.
\end{equation}
Because $\zt<\zn$, the values of $D_{mm}$ can be bounded as $-3\leq
D_{mm}\leq 1$.

Equation \eqref{curv_dev_simp:eq} contains one additional term
involving the second derivative that is not present in the
corresponding low amplitude equation \eqref{curveq}.  Below we argue
that the additional term is small even when $\kappa_{0}$ itself is not
small. Our analysis in the main text relied on performing an
eigenfunction expansion using the eigenfunctions of the beam equation.
We use the same expansion here
\begin{equation}
  c = \sum_{k=1}^{\infty} \alpha_{k}(t)\Psi_k(s),
\end{equation}
and we relate $c$ to its expansion coefficients by
\begin{equation}
  c(s,t) = P^{-1} \vec{\alpha}(t),
\end{equation}
where $P^{-1}$ is the orthogonal operator that maps the expansion
coefficients to $c$. We can transform equation
\eqref{curv_dev_simp:eq} into a system of differential equation for
the expansion coefficients as
\begin{equation}
 \frac{d\vec{\alpha}}{dt} + P\partial_{t}\kappa_{0} =
   -\frac{\kbdim}{\zn} N^{4}\vec{\alpha} +
   \frac{\kbdim\kappa_{0}^{2}}{\zt}P Q^{-1}DQ P'' \vec{\alpha}
\end{equation}
%
%\begin{align}
%  \frac{d\vec{\alpha}}{dt} + P\partial_{t}\kappa_{0} & =
%  -\frac{B}{\zn} \Lambda^{4}\vec{\alpha} +
%  \frac{B\kappa_{0}^{2}}{\zt} P Q^{-1}DQ  c_{ss} \\
%  &  =  -\frac{B}{\zn} \Lambda^{4}\left( I +
%  \frac{\zn\kappa_{0}^{2}}{\zt} \Lambda^{-4}P Q^{-1}DQ P''\right) \vec{\alpha},
%\end{align}
%
where we
define $P''$ so that $P''\vec{\alpha}= c_{ss}$, and 
$-N^{4}$ is a diagonal matrix containing the
eigenvalues of the beam equation. That is, the $k^{th}$ diagonal entry of
$N$, $\nu_{k}$, is related to the  $k^{th}$ eigenvalue, $\mu_{k}$, by $\mu_{k}=-\nu_{k}^{4}$. One expects that the
contribution of $P''$ to the $k^{th}$ equation to scale like
$\nu^{2}_{k}$. As argued above the norm of $Q^{-1}DQ$ is about 1,
and so we expect the additional terms relative to the bending terms to
contribute $\nu_{k}^{-2}$. The smallest eigenvalue is about
$\nu^{4}_{1}\approx 500$, and thus we expect these additional
terms to be small.

\begin{figure}
  \centering
  \includegraphics[width=0.95\textwidth]{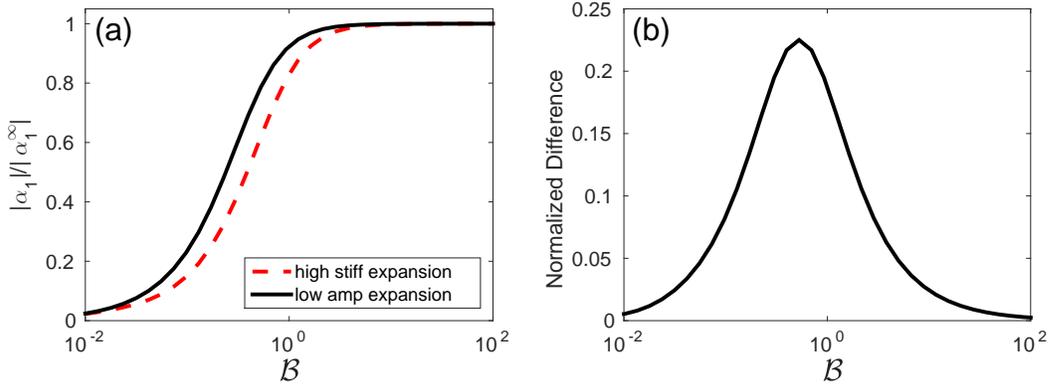}
  \caption{(a) Normalized amplitude of first mode for a
    high-amplitude flexor with amplitude $\kappa_{0}=4$ as a function
    of stiffness $\kbdim$ for the leading order low amplitude expansion and the
    leading order high stiffness expansion. The ratio of drag was set
    to $\zt/\zn=0.75$ for this computation. The results are relatively
    insensitive to this value. (b) Magnitude of the difference of
    the two expansion coefficients.}
  \label{expans_comp:fig}
\end{figure}

In figure \ref{expans_comp:fig} we compare the expansion coefficients
of the first mode of the small amplitude expansion and the high
stiffness expansion for the high-amplitude flexor ($\kappa_{0}=4$) for
a range of stiffnesses. In agreement with our numerical results, the
qualitative behavior of the two expansions is the same, and at high
and low stiffnesses the quantitative behavior is the same. The largest
difference occurs for moderately soft bodies where the difference is
less than 25\%.

%%%%%%% Appendix 2

\section{Rod driven at one end}\label{App:enddriven}

In this appendix we give the derivation for the shape-induced swimming speed of 
a thin filament in a viscous or viscoelastic fluid which is driven by oscillations at one end. 
This type of motion is akin to the experiments of \cite{espinosa2013fluid}, and we perform
the calculation to compute the swimming speed as a function of the dimensionless body response time
$\G$ and the fluid relaxation time $\De.$ We also show the range of $\G$ where a viscoelastic
 speed-up is theoretically predicted for this type of motion.

\subsection{Shape of the swimmer}

The problem we consider is a flexible filament with one free end and one 
clamped end. The motion is driven by prescribing sinusoidal oscillations in the
angle at the clamped end.  The
shape of the filament satisfies
\begin{gather}
  y_{t} =  -G y_{ssss} \\
  y(0,t) = 0 \\ y_{s}(0,t) = \cos(2\pi t) \\
  y_{ss}(1,t) = 0  \\ y_{sss}(1,t) = 0.
\end{gather}
We now change variables using
\begin{equation}
  y = w + s\cos(2\pi t)=w+\textrm{Re}(s\textrm{e}^{2\pi i t}),
  \label{ywrel:eq}
\end{equation}
so that $w$ represents deviations from the infinitely stiff case of a
straight rod. Letting $w$ be complex valued, the equation for $w$ is
then
\begin{equation}
    w_{t} =  -G w_{ssss}  - 2\pi i s\textrm{e}^{2\pi i t}
\end{equation}
with homogeneous boundary conditions. This equation can be solved
using an expansion of eigenfunctions, $\Psi_{k}(s)$, which satisfy
\begin{gather*}
  \mu \Psi(s)=-\Psi_{ssss},\\
  \Psi(0)=\Psi'(0) = 0, \\
  \Psi''(1)=\Psi'''(1) = 0.
\end{gather*}
We express the function $s$ using an eigenfunction expansion
\begin{equation}
  s= \sum_{k} \alpha_{k}^{\infty} \Psi_{k}(s),
\end{equation}
and look for a solution to the PDE  of the form
\begin{equation}
  w(s,t) = \textrm{e}^{2\pi i t} \sum_{k}\beta_{k} \Psi_{k}(s).
\end{equation}
The transformation of the PDE yields
\begin{equation}
  2\pi i \beta_{k} = G\mu_k \beta_{k} - 2\pi i \alpha_{k}^{\infty},
\end{equation}
which gives
\begin{equation}
  \beta_{k} = \frac{-\alpha^{\infty}_{k}}{1 - \frac{G\mu_k}{2\pi i}}.
\end{equation}
We can then write the shape as
\begin{align}
  y(s,t) & = Re\left\{ \sum_{k=1}^{\infty}(\alpha_{k}^{\infty}+\beta_{k})\Psi_{k}(s)\textrm{e}^{2\pi i t} \right\}, \\
  & = Re\left\{ \sum_{k=1}^{\infty}\alpha_{k}^{\infty}\left(1 -\left(1 - \frac{G\mu_{k}}{2\pi i}\right)^{-1} \right)\Psi_{k}(s)\textrm{e}^{2\pi i t} \right\}.
\end{align}
Notice that the factor multiplying $\alpha_{k}^{\infty}$ above is
exactly the same as the one that appears in the shape analysis of
swimmers driven by active moments in \eqref{visc_curv}. We can express
the shape as 
\begin{equation}
 y(s,t) = Re\left\{ \sum_{k=1}^{\infty}\alpha_{k} \Psi_{k}(s)\textrm{e}^{2\pi
   i t} \right\},
\end{equation}
where $\alpha_{k}$ is defined by \eqref{visc_curv}. As in the main
body of the paper, to add viscoelastic effects, we simply use
\eqref{veshapechanges} in place of \eqref{visc_curv} to define the
expansion coefficients. Although these expressions are the same, we
note that the eigenfunctions and eigenvalues are different for this
problem.

\subsection{Expression for Swimming Speed}
The expansion for the shape of the swimmer can be written as
 \begin{equation}
  y(s,t) = \sum_{k}^{\infty} A_{k}\cos(2\pi t + \phi_{k})\Psi_{k}(s),
\end{equation}
 where
\begin{gather}
  A_{k} = \left| \alpha_{k} \right| \\
  \phi_{k}   = \textrm{arg}(\alpha_{k}).
\end{gather}
To compute the swimming speed, we average in space and time the
product
\begin{equation}
  y_{s}y_{t} = -2\pi \sum_{n}\sum_{m} A_{n}A_{m} \cos(2\pi t + \phi_{n})\sin(2\pi t + \phi_{m})\Psi_{n}(s)\Psi'_{m}(s).
\end{equation}
Averaging the above expression by integration gives the swimming speed
\begin{equation}\label{app:ss}
  \left<{U}\right>  \propto
   \sum_{n}\sum_{m} A_{n}A_{m}\sin(\phi_{n}-\phi_{m})\int_{0}^{1}\Psi_{n}(s)\Psi'_{m}(s)\,ds.
\end{equation}

\begin{figure}
  \centering
 \includegraphics[width=0.95\textwidth]{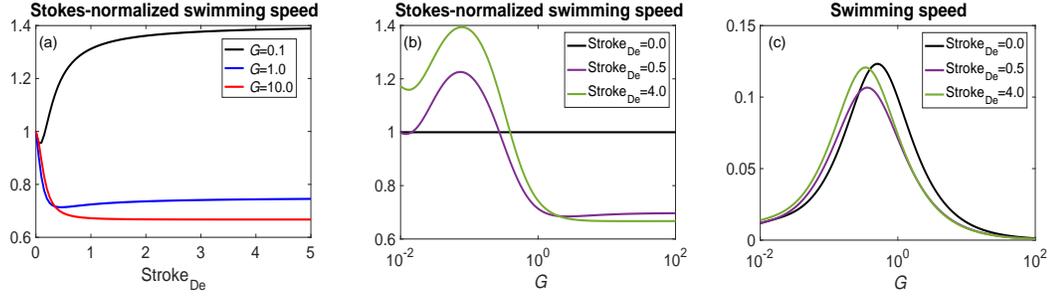}
  \caption{(a) Shape induced Stokes-normalized swimming speed as a function of $\strDe$ for a range of $\G$. 
  (b) Shape induced Stokes-normalized swimming speed as a function of $\G$ for low and high $\strDe$. (c) Shape induced swimming speed as a function of $\G$ for low and high $\strDe$.}
  \label{cantilever:fig}
\end{figure}

We use this expression with the first six modes to compute the
shape-induced Stokes-normalized swimming speed; see figure
\ref{cantilever:fig}. As with the problem from the paper, for
sufficiently soft bodies, we see an almost monotonic speed-up from the
shape changes. For sufficiently stiff swimmers ($\G> 1$) we see a
monotonic slow down. There is a transition range around
$0.1<\G<1$. The location of this transition is evident in figure
\ref{cantilever:fig} (b) where we show the swimming speed as a
function of $\G$ for low and high $\strDe$. The qualitative results from the paper do not change in the sufficiently soft regime, but this problem
has a different driving mechanism and hence there is a different effect in the stiff regime. As we see in figure
\ref{cantilever:fig} (c) as the body is stiffened the swimming speed goes to zero and viscoelasticity always slows the 
swimmer. 

\bibliographystyle{unsrt}
\bibliography{flex_bib_BT}

\end{document}